\documentclass[shortnote,twocolumn]{jpsj2} 
\title{%
Note on Spin Structure of the Classical Vector Spin Heisenberg Model
}%
\author{%
Kiyosi \textsc{ Terao}\thanks{E-mail address: terk005@shinshu-u.ac.jp}
}%
\inst{%
Faculty of Science, Shinshu University, Matsumoto 390-8621
}%

\recdate{ May 4, 2005 }

\kword{%
spin structure, classical spin, Heisenberg model, chirality
}%
\begin{document}
\maketitle

The energy of  the classical spin Heisenberg model is written as
\begin{equation}
	E = -\frac{1}{2} {\sum_{\mib{n}\nu, \mib{m}\mu}}
	\,J_{\mib{n}\nu, \mib{m}\mu}
	\mib{s}_{\mib{n}\nu} \cdot \mib{s}_{\mib{m}\mu},
\end{equation}
where $\mib{s}_{\mib{n}\nu}$ is the spin at the $\nu$-th site in the $\mib{n}$-th unit cell and $J_{\mib{n}\nu, \mib{m}\mu}$ is
the exchange parameter connecting spins at $\mib{n}\nu$ and $\mib{m}\mu$ atoms. 
We treat the spin $\mib{s}_{\mib{n}\nu}$ as a classical vector of magnitude $s$. The energy is written using Fourier components as
\begin{equation}
	E= -\frac{N}{2} \sum_{\mib{k}} \sum_{\nu,\mu}
	J_{\nu \mu}(\mib{k})
	\mib{s}_{\nu}(\mib{k}) \cdot \mib{s}_{\mu}(-\mib{k}),
		\label{E_k}
\end{equation}
where $\mib{k}$ is the wave vector in the first Brillouin zone and
\begin{align}
	\mib{s}_{\mib{n}\nu}= \sum_{\mib{k}} 
			\mib{s}_{\nu} (\mib{k}) \exp ({\rm i} \mib{k} \cdot \mib{R}_{\mib{n}\nu}), 
	\ \ \nu=1,2,\cdots,p,
\end{align}
and
\begin{equation}
J_{\nu\mu}(\mib{k}) = \sum_{\mib{n}}
			J_{\mib{n}\nu,\mib{m}\mu}
			\exp [{\rm i} \mib{k} \cdot (\mib{R}_{\mib{n}\nu}-\mib{R}_{\mib{m}\mu})].
\end{equation}
The ground-state spin structure is determined by minimizing $E$ in eq. (\ref{E_k}) under the constraint
\begin{align}
	\mib{s}_{\mib{n}\nu} \cdot \mib{s}_{\mib{n}\nu} &= \sum_{\mib{k}, \mib{k}'}
	\mib{s}_{\nu}(\mib{k}) \cdot \mib{s}_{\nu}(\mib{k}')\exp[{\rm i}(\mib{k}+\mib{k}')\cdot \mib{R}_{\mib{n}\nu}]\notag \\
    &= s^2
		\label{const_k}
\end{align}
for any $\mib{R}_{\mib{n} \nu}$, from which
\begin{subequations}
\begin{align}
& \sum_{\mib{k}} \mib{s}_{\nu}(\mib{k}) \cdot \mib{s}_{\nu}(-\mib{k}) = s^2, \\
& \sum_{k}\mib{s}_{\nu}(\mib{k}) \cdot \mib{s}_{\nu}(\mib{k}')
	= 0 \ \ \ \mbox{for} \ \ \mib{k}+\mib{k}' \ne 0.
\end{align}
\end{subequations}
If we assume that the ground state is described by a single pair of $\mib{k}= \mib{q} \mbox{ and } -\mib{q}$,  i.e.,
\begin{equation}
\mib{s}_{\mib{n}\nu}= \mib{s}_{\nu} (\mib{q}) \exp ({\rm i} \mib{q} \cdot \mib{R}_{\mib{n}\nu})
+\mib{s}_{\nu} (\mib{-q}) \exp (-{\rm i} \mib{q} \cdot \mib{R}_{\mib{n}\nu}),
			\label{single_q}
\end{equation}
 the constraint condition is written as
\begin{subequations}
  \begin{align}
	&2\mib{s}_{\nu}(\mib{q}) \cdot \mib{s}_{\nu}(-\mib{q}) = s^2, \label{const_s}\\
	&\mib{s}_{\nu}(\mib{q}) \cdot \mib{s}_{\nu}(\mib{q}) = 0.
		\label{const_0}
  \end{align}
\end{subequations}
On the basis of this assumption, Yoshimori\@~\cite{YA} and Nagamiya\@~\cite{N} have studied the helical (screw) spin structure. Hereafter, we refer to this assumption as the single $\mib{q}$ model.
Decomposing $\mib{s}_{\nu}(\mib{q})$ into real and imaginary parts, $\mib{s}_{\nu}(\mib{q}) = \mib{s}_{\nu}'(\mib{q})   + \mathrm{i} \mib{s}_{\nu}''(\mib{q})$, we may rewrite the constraint condition, eqs. (\ref{const_s}) and (\ref{const_0}), as
\begin{subequations}
\begin{align}
	&2[\mib{s}_{\nu}'(\mib{q})^2 + \mib{s}_{\nu}''(\mib{q})^2] = s^2, \\
	&\mib{s}_{\nu}'(\mib{q})^2 - \mib{s}_{\nu}''(\mib{q})^2 =0, \\ 
	&\mib{s}_{\nu}'(\mib{q}) \cdot \mib{s}_{\nu}''(\mib{q}) =0.
\end{align}
\end{subequations}
As $\mib{s}_{\nu}'(\mib{q})$ is orthogonal to $\mib{s}_{\nu}''(\mib{q})$,   $\mib{s}_\nu(\mib{q})$ is represented as
\begin{equation}
	\mib{s}_\nu(\mib{q})
	=\frac{s}{2}u_\nu(\mib{q})[\mib{i}_{\nu}(\mib{q})-{\rm i}\mib{j}_{\nu}(\mib{q})],
		\label{sij}
\end{equation}
using a set of orthonormal real vectors $\mib{i}_\nu(\mib{q})$ and $\mib{j}_\nu(\mib{q})$,
where $u_\nu(\mib{q})$ is a phase factor. For two sites per unit cell, Yoshimori\@~\cite{YA} and Nagamiya\@~\cite{N} have pointed out that [$\mib{i}_1(\mib{q})$, $\mib{j}_1(\mib{q})$] and [$\mib{i}_2(\mib{q})$, $\mib{j}_2(\mib{q})$] lie on the same plane.  
In this note, we prove that [$\mib{i}_\nu(\mib{q})$, $\mib{j}_\nu(\mib{q})$] are independent of $\nu$ for any number of sites per unit cell. 

Now, we consider the translational invariance of $\mib{s}_{\mib{n}\nu} \cdot \mib{s}_{\mib{m}\mu}$ for $\mib{R}_{\mib{n}\nu} \ne \mib{R}_{\mib{m}\mu}$. For the single $\mib{q}$ model, we have 
\begin{align}
	& \mib{s}_{\mib{n}\nu} \cdot \mib{s}_{\mib{m}\mu}
	= \ \mib{s}_\nu (\mib{q}) \cdot \mib{s}_\mu (\mib{q}) 
	\exp[{\rm i} \mib{q} \cdot (\mib{R}_{\mib{n}\nu}+\mib{R}_{\mib{m}\mu})] \notag \\
	& + \mib{s}_\nu (\mib{q}) \cdot \mib{s}_\mu (-\mib{q})
	\exp[{\rm i} \mib{q} \cdot (\mib{R}_{\mib{n}\nu}-\mib{R}_{\mib{m}\mu})] \notag \\
	& + \ \ {\rm complex\  conj.}
\end{align}
This should be independent of $\mib{R}_{\mib{n}\nu}+\mib{R}_{\mib{m}\mu}$ because of the translational invariance of $E$. Then, 
\begin{equation}
\mib{s}_\nu (\mib{q}) \cdot \mib{s}_\mu (\mib{q}) =0
\end{equation}
for any $\nu$ and $\mu$. Using eq.(\ref{sij}),
\begin{align}
	& \mib{s}_\nu (\mib{q}) \cdot \mib{s}_\mu (\mib{q}) \notag \\
	&= \frac{s^2}{4} u_\nu (\mib{q}) u_\mu (\mib{q})
	\Bigl\{ \bigl[ \mib{i}_\nu (\mib{q}) \cdot \mib{i}_\mu (\mib{q})
	-\mib{j}_\nu (\mib{q}) \cdot \mib{j}_\mu (\mib{q}) \bigr] \notag \\
	&\  -{\rm i}  \bigl[ \mib{i}_\nu (\mib{q}) \cdot \mib{j}_\mu (\mib{q})
	-\mib{j}_\nu (\mib{q}) \cdot \mib{i}_\mu (\mib{q}) \bigl] \Big\}.
\end{align}
Thus, we obtain
\begin{subequations}
\begin{align}
	\mib{i}_\nu (\mib{q}) \cdot \mib{i}_\mu (\mib{q})
	-\mib{j}_\nu (\mib{q}) \cdot \mib{j}_\mu (\mib{q}) &= 0, \label{newcond1}\\
	\mib{i}_\nu (\mib{q}) \cdot \mib{j}_\mu (\mib{q})
	-\mib{j}_\nu (\mib{q}) \cdot \mib{i}_\mu (\mib{q}) &=0.	\label{newcond2}
\end{align}
\end{subequations}
Without loosing generality, we set  an $(x, y, z)$-coordinate system as $\mib{i}_\nu (\mib{q})=(1, 0, 0)$ and $\mib{j}_\nu (\mib{q})=(0, 1, 0)$.
We assume $[ \mib{i}_\mu (\mib{q}), \mib{j}_\mu (\mib{q}) ]$ is obtained from $[ \mib{i}_\nu (\mib{q}), \mib{j}_\nu (\mib{q}) ]$ by rotating the Eulerian angles $\psi$ about the $z$-axis, $\theta$ about the $y'$-axis and the $\phi$ about the $z''$-axis, i.e.,
\begin{subequations}
\begin{align}
	&\mathcal{ R}(\psi, \theta, \phi) \mib{i}_\nu(\mib{q}) = \mib{i}_\mu(\mib{q}) \notag \\
	= & \begin{pmatrix}
	\cos \psi \cos \theta \cos \phi -\sin \psi \sin \phi \\
	\sin \psi \cos \theta \cos \phi + \cos \psi \sin \phi \\
	-\sin \theta \cos \phi
	\end{pmatrix}, \\
	&\mathcal{ R}(\psi, \theta, \phi) \mib{j}_\nu(\mib{q}) = \mib{j}_\mu(\mib{q}) \notag \\
	= & \begin{pmatrix}
	-\cos \psi \cos \theta \sin \phi -\sin \psi \cos \phi \\
	-\sin \psi \cos \theta \sin \phi + \cos \psi \cos \phi \\
	\sin \theta \sin \phi
	\end{pmatrix},
\end{align}
\end{subequations}
where $\mathcal{ R}(\psi, \theta, \phi)$ denotes the rotation operator. 
Then eqs. (\ref{newcond1}) and (\ref{newcond2}), respectively, are represented by the Eulerian angles as  
\begin{subequations}
\begin{align}
	&\cos(\psi-\phi)[\cos \theta -1] = 0, \\
	&\sin(\psi-\phi)[\cos \theta -1] = 0.
\end{align}
\end{subequations}
Consequently, $\theta = 0$, 
which means that the rotation consistent with the translational invariance of $E$ is the one about the $z$-axis by angle $\psi + \phi$. The orthonormal sets $[ \mib{i}_\nu (\mib{q}), \mib{j}_\nu (\mib{q})]$ and $[\mib{i}_\mu (\mib{q}), \mib{j}_\mu (\mib{q})]$ are on parallel planes. The rotation by angle $\psi + \phi$ about the $z$-axis can be included in the initial phase difference between $u_{\nu}(\mib{q})$ and $u_{\mu}(\mib{q})$. Then we can choose an orthonormal set [$\mib{i}(\mib{q})$, $\mib{j}(\mib{q})$] without considering the sites. The spins at different sites rotate in parallel [$\mib{i}(\mib{q})$, $\mib{j}(\mib{q})$] planes. The orientation of the planes is not related to the crystal axis or the direction of $\mib{q}$.
     
Thus, we have resolved the ambiguity about the site dependence of [$\mib{i}(\mib{q})$, $\mib{j}(\mib{q})$] in the review by Nagamiya.~\@\cite{N} 
Note that the single-$\mib{q}$ assumption, eq.(\ref{single_q}), excludes the chiral order parameter of the vector triple product, $\vec{\chi} =\mib{s}_i \cdot [\mib{s}_j \times \mib{s}_k]$, of any spins.  It is intriguing whether $\vec{\chi}$ can appear as an order parameter in a translationally ordered ground state.

\end{document}